\documentstyle[12pt]{article}
\begin{document}
\begin{titlepage}

\begin{flushright}
PSI-PR-95-4\\
\end{flushright}
\begin{flushright}
January, 1995\\
\end{flushright}

\vspace{2cm}
\begin{center}

\begin{large}
{\bf Spontaneous breaking of  vector symmetries
and the non-decoupling light Higgs particle}
\end{large}

\vspace*{2cm}
        {\bf Hanqing Zheng$^\dagger$ }\\
\vspace*{1.0cm}
        Paul Scherrer Institute, 5232 Villigen PSI, Switzerland

\end{center}
\vspace*{2cm}
\begin{abstract}
Using  a four fermion interaction Lagrangian,
we demonstrate that the spontaneous breaking of vector symmetries requires
the existence of a light (comparing with the heavy fermion
mass) scalar particle and
the low energy effective theory (the $\sigma$
model) obtained after integrating out heavy fermion
degrees of freedom is asymptotically
a renormalizable one. When applying
the idea to the electroweak symmetry breaking sector of
the standard model, the Higgs particle's mass is of the order of the
electroweak scale.
\end{abstract}

\vfill
\vfill

\begin{flushleft}
$\dagger$) e-mail address: zheng@cvax.psi.ch
\end{flushleft}

PACS Number(s): 11.30.Qc, 12.60.Fr

\phantom{p}
\vfill

\phantom{p}
\end{titlepage}

The  spontaneous breaking of vector symmetries (SBVS)
is  an interesting subject in quantum field theory
and also in particle physics, as long as it remains to be
a possibility as a character of nature.
In a previous paper~\cite{1} we have made an attempt in
considering
the breaking of the electroweak symmetry
as a consequence of the SBVS between  fermions with heavy bare
masses. The
motivation of such a consideration is to break the electroweak symmetry
dynamically but with least influence on the low energy physics \cite{1}.
We modeled  SBVS by a low energy effective Higgs--Yukawa interaction
and, after integrating out the heavy fermion fields in the mean field
approximation of the Higgs particle, estimated the low
energy residual effects of these heavy fermions.
It is shown that the heavy fermion fields are essentially decoupling
at low energies, except that they can
generate massless Goldstone excitations
to be absorbed by the weak gauge fields.
The low energy effective theory (the standard model) should therefore,
be weakly interacting -- a picture different from the
technicolor models.

Nothing can be said about the mass of the composite
Higgs particle, within the
context of the effective Higgs--Yukawa model in ref.~\cite{1}.
A question then arise: Can it be
as heavy as the heavy fermion mass? From
general physical consideration we know that this should
not be the case. Because
if the Higgs particle's mass is  heavy enough, the remaining fields must
be in a strongly interacting system because of the well known
tree--level
unitarity argument \cite{chanowitz} -- a result contrary to our
motivation and the general expectation from the
decoupling phenomena. The aim of this paper is to resolve this Higgs mass
ambiguity,
using a model of four fermion interactions with the dynamically
generated SBVS.
The SBVS through four fermion interactions was first
analysed by Preskill and Weinberg \cite{preskill}
to study the possible violation of the
"persistent mass condition".
For a four fermion
interaction with a global vector symmetry, there are primarily
two scales, the cutoff scale $\Lambda$ and the bare
fermion mass $M$($M<\Lambda$), and in addition, the
interaction strength is characterized by a dimensionless
coupling constant, $G$. Preskill and Weinberg have shown
that, for a given cutoff $\Lambda$ and a sufficiently large $G$, there
exists a critical value $M_c$. When the fermion mass is below
this critical point, $M<M_c$, the vector (isospin) symmetry is
spontaneously broken down. As a consequence, there exist
massless particles composed
of massive constituents leading to a violation of the persistent mass
condition. When $M$ exceeds $M_c$, the system is in symmetry phase and
the decoupling phenomenon occurs.
The symmetry breaking is of second order and
characterized by a new scale $m$ (the fermion mass
splitting) obtained after some fine tuning.

For our purpose, the four fermion interaction Lagrangian can be
written as\footnote{One can add more terms and couplings, see
for example ref.~\cite{1}. The present
Lagrangian is the minimal one suitable for the discussion in the present
paper.}
\begin{equation}
\label{4f}
{\cal L}=\bar\Psi^i \left ( i\partial\llap{$/$}-M \right ) \Psi^i
-{G\over N_c\Lambda^2}[ (\bar \Psi^i\rho_3\Psi^i)^2+
(\bar \Psi^i\rho_1\vec\tau\Psi^i)^2 ]\ ,
\end{equation}
where
$\Psi = (\psi_1,\psi_2)^T$ and  $\psi_{1,2}$ are SU(2)
isospin doublets. The index i refers to the "color" degree of freedom
and runs over 1 to $N_c$. We assume $N_c$ is large in the following.
$\tau_i$ are generators
of the  SU(2) isospin group
and $\rho_i$ are Pauli matrices of the "parity doublet" space
(i.e., space between $\psi_1$ and $\psi_2$). The Lagrangian equation
(\ref{4f}) is invariant under the following  $SU(2)\times
SU(2)$ rotations:
\begin{equation}
\Psi \rightarrow e^{i\vec\alpha\cdot\vec\tau +
      i\rho_2\vec\beta\cdot\vec\tau}\Psi \ .
\end{equation}

To match the electroweak physics one of the $SU(2)$ global symmetry will
be gaugeized as $SU(2)_W$ (of course, the local $U(1)_Y$ should also
be  introduced). The another "custodial" $SU(2)$ symmetry remains
as a global one and can be broken explicitly but  slightly.
The latter constraint
comes from the experimental value of the $\rho$ parameter.
Since these
 are already discussed  in ref.~\cite{1} and
are not  very relevant
to the topic in the present paper we will no longer
discuss them but just work with the
effective Lagrangian eq.~(\ref{4f}).
To study the spontaneous symmetry breaking we look at the
gap equation
and search for a solution of the fermion
mass matrix of the type $m=m_s+\rho_3m_3$. In the large $N_c$ limit
we obtain,
\begin{eqnarray}
m_s & = & M \ ,\\
m_3 & = & {iG\over \Lambda^2}\int^\Lambda{d^4p\over (2\pi)^4}
tr(\rho_3S_F) \ , \label{gap}
\end{eqnarray}
where,
\begin{equation}
S_F=\left(\matrix{S_F^1 &  \cr  & S_F^2 \cr}\right)\ .
\end{equation}
The above equation (\ref{gap}) can be written in a simple form. To
define
\begin{equation}
f(m) = {m\over \Lambda^2}\int^\Lambda{q_E^2dq_E^2\over q_E^2+m^2}\ ,
\end{equation}
we have,
\begin{equation}
m_1-m_2= {G\over \pi^2}[f(m_1)-f(m_2)]\ ,\end{equation}
where $m_{1,2}=M\pm m_3$. For small but non-vanishing $m_1-m_2$
the above equation can be further approximated as
\begin{equation} \label{gap'}
{\pi^2\over G}\simeq f'(M) + 1/6 m_3^2 f'''(M)\ .
\end{equation}
For small values of $M$ (or $M_c$)\footnote{In
the present case, the gap
equation  is  more sensitive than in the Nambu--Jona-Lasinio (NJL)
model
to the higher order terms in the $1/\Lambda$ expansion. Unambiguous
results can only be obtained when keeping $M/\Lambda $ small,
since these
subleading terms are regularization scheme dependent.},
$f'(M)$ is a decreasing
function of $M$ and $f'''(M)$
is negative. Therefore we observe from the above formula that when
$\pi^2/G$ is smaller than unity there exists the critical value $M_c$,
$\pi^2/G=f'(M_c)$. When $M$ is less than $M_c$ there exists a
non-vanishing solution of $m_3$,
\begin{equation}
m_3=\sqrt{6M_c(M_c-M)}\ ,
\end{equation}
which holds in the $m_3<<M$ (or $M\rightarrow M_c$) limit.
Once $M$ exceeds the critical
value $M_c$ there is only the trivial solution $m_3=0$ in the above
equation (\ref{gap'}).

Up to now we have said little more  than the result obtained
in ref.~\cite{preskill} except that in our case the vector group
is $SU(2)\times SU(2)$ which spontaneously breaks down to $SU(2)$
and therefore there are 3 Goldstone bosons. The appearance of
these massless Goldstone excitations
implies that the decoupling
of the heavy fermions realizes in a nontrivial manner because of the
existence of the phase transition.

To understand more about the dynamics of  SBVS it is necessary to
solve the Lagrangian eq.~(\ref{4f}) in the large $N_c$ limit. For
our purpose it is appropriate to discuss
the following two point functions,
\begin{equation}
\Pi_P(q^2) \equiv i\int d^4x e^{iqx}<|T\{\bar \Psi^i(x)
\rho_1\Psi^j(x)\bar\Psi^j(0)\rho_1\Psi^i(0)\}|>\ ,\end{equation}
\begin{equation}
\Pi_S(q^2) \equiv i\int d^4x e^{iqx}<|T\{\bar \Psi(x)
\rho_3\Psi(x)\bar\Psi(0)\rho_3\Psi(0)\}|>\ ,
\end{equation}
\begin{eqnarray*}
\Pi_M^\mu(q^2) &\equiv& i\int d^4x e^{iqx}<|T\{\bar \Psi^i(x)
\rho_2\gamma^\mu\Psi^j(x)\bar\Psi^j(0)\rho_1\Psi^i(0)\}|>
\end{eqnarray*}
\begin{equation}
\equiv iq^\mu\Pi_M(q^2)\ .
\end{equation}
In above equations $i,j$ denote isospin indices and we dropped out the
color indices for simplicity. These two point functions are obtainable
by summing up  fermion bubble chains. We use the overlined functions
to denote the 1--loop contribution to the two point functions.
Direct calculation leads to,
\begin{equation}\label{P}
\Pi_P(q^2)={\overline \Pi_P(q^2)\over 1-G/\Lambda^2
\overline \Pi_P(q^2)}\ ,
\end{equation}
\begin{equation}\label{S}
\Pi_S(q^2)={\overline \Pi_S(q^2)\over 1-G/\Lambda^2
\overline \Pi_S(q^2)}\ ,
\end{equation}
and
\begin{equation}\label{M}
\Pi_M^\mu(q)={\overline \Pi_M^\mu(q)\over 1-G/\Lambda^2
\overline \Pi_P(q^2)}\ ,
\end{equation}
where  $\overline \Pi_P$ and $\overline \Pi_S$
are quadratically divergent and $\overline \Pi_M^\mu$
only contains logarithmic divergence
(the latter one is linearly divergent in NJL model).
One must be careful in dealing with the quadratic divergence
in order to avoid the dependence on the choice of the internal momentum
flow. The standard method to overcome this difficulty is to calculate
firstly
the imaginary part of the two point functions using Cutkosky rule
and then use dispersion
relations to evaluate the full amplitudes. The dispersion integrals
are usually divergent and need subtractions.
To deal with $\overline
\Pi_P$ it is useful to rewrite it as
$\overline \Pi_P(q^2)=\overline \Pi_P(0) + q^2\overline
\Pi'_P(q^2)$. The
function $\overline \Pi'_P(q^2)$ (which coincides with
$d/dq^2 \overline \Pi_P(q^2)$ at origin) now only contains logarithmic
divergence, the quadratic cutoff dependent term is already
absorbed into $\overline \Pi_P(0)$. In order to have a Goldstone pole
in the function $\Pi_P(q^2)$ we read off the self-consistency condition
for  SBVS which should be equivalent to the gap equation,
\begin{equation}\label{sc}
1-G/\Lambda^2\overline \Pi_P(0)=0\ .
\end{equation}
The mixed function $\Pi_M$ can then be written as,
\begin{equation}\label{pM}
\Pi_M={\overline \Pi_M(q^2)\over -G/\Lambda^2 q^2 \overline
\Pi'_P(q^2)}\ ,
\end{equation}
while\footnote{
$
{1\over \pi}\hbox{Im} \overline \Pi_P(t) =
{1 \over 4\pi^2} (t-(m_1+m_2)^2)\sqrt{(1-{(m_1+m_2)^2\over t})
(1-{(m_1-m_2)^2\over t})},$
$
{1\over \pi}\hbox{Im} \overline \Pi_S(t) ={1\over 2}\{
{1 \over 4\pi^2} (t-4m_1^2)\sqrt{1-{4m_1^2\over t}}+
(m_1\to m_2)\}$. At the critical point of the phase transition
($m_1=m_2$) these two
functions are equal. This is of course the consequence of the symmetry.}
\begin{eqnarray*}
{1\over \pi}\hbox{Im} \overline \Pi_M(t) &=&
{m_1-m_2\over 4\pi^2} (1-{(m_1+m_2)^2\over t})^{3/2}
(1-{(m_1-m_2)^2\over t})^{1/2} \\
\end{eqnarray*}
\begin{equation}
=(m_1-m_2) {1\over \pi}\hbox{Im}\overline \Pi'_P(t) \ .
\end{equation}
The use of unsubtracted dispersion relations (with a truncated integrand
at $4\Lambda^2$)
as proposed in ref.~\cite{3} immediately leads to
$\overline\Pi_M/\overline \Pi_P' \equiv const$. This is not  an accident,
as can be proven using the equal time anti-commutation relation of the
quark fields and the current conservation condition\footnote{
Similar results were obtained in NJL model in ref.~\cite{2} in where
different regularization schemes are used.},
\begin{equation}
\Pi_M^\mu \equiv {2iq^\mu\over q^2}<|\overline\Psi\rho_3\Psi|>\ .
\end{equation}
Therefore from eq.~(\ref{pM}) we obtain,
\begin{equation}
<|\overline \Psi\rho_3\Psi |>= -{N_c\over 2G}\Lambda^2(m_1-m_2)\ .
\end{equation}
For the  two point function $\overline\Pi_S(q^2)$,
one can write $\overline\Pi_S(q^2)=\overline \Pi_P(q^2) +
\delta\overline\Pi_S(q^2)$.
Again the quadratic divergence is absorbed into
$\overline\Pi_P(0)$ and $\delta\overline\Pi_S(q^2)$ only contains
logarithmic divergence. The fine tuning
is isolated in eq.(\ref{sc}) or in the gap equation.
The Higgs particle's mass is obtained by looking for the pole position
of the scalar two point function.
We read off from eq.~(\ref{S}) that,
\begin{equation}
m_H^2=-\delta\overline\Pi_S(m_H^2)/\overline\Pi'_P(m_H^2)\ .
\end{equation}
In the symmetry phase (i.e., $m_1=m_2$)
$\Pi_S$ is identical to $\Pi_P$ as a result of the global symmetry.
Since these Green functions are continuous in $m_3$,
$\delta\overline\Pi_S\sim (m_1-m_2)^2$ and therefore $m_H^2$ is a small
quantity. Approximately we have
$m_H^2= -\delta\overline\Pi_S(0)/\overline\Pi'_P(0)$. Simple
calculation
yields\footnote{
This expression receives $O(1/\ln(\Lambda/M))$
corrections which can not be determined unambiguously~\cite{hasen},
although it is practically unimportant in the present case.
Especially adding more four fermion interaction terms with higher
derivatives
in the effective Lagrangian as pointed out in the first paper of
ref.~\cite{hasen} will not lead $m_H$ to be proportional
to $M$ rather than $m_3$.},
\begin{equation}
m_H=2m_3\ .
\end{equation}
This result indicates that the scalar  particle's mass
is small, i.e., at the symmetry breaking scale
(comparing with the  fermion mass scale and the cutoff
parameter).  Especially it has
nothing to do with the whole fermion mass,
rather it is only related to the dynamically generated part of the
fermion mass. In the
NJL model for chiral symmetry breaking it happens to be
that the two masses coincide.
It is worth pointing out that
the light mass of the scalar particle composed of heavy
fermion fields  is a
 consequence of the symmetry and be model independent, at least
in a system with second order phase transition.
This property is not
shared by other possible composite particles.
For example one could add
another four fermion interaction term in the Lagrangian  eq.(\ref{4f})
with vector--vector couplings and the mass of the
vector resonance, if exists, is $\sim \Lambda$ and can be
large.

To discuss the electroweak physics, comparing with
the expression of the
decay constant of the Goldstone field \cite{1}, $f_\pi^2=
{N_c\over 2\pi^2} m_3^2\ln(\Lambda^2/M^2)$, we obtain,
\begin{equation}
m_H={2\pi v\over \sqrt{N_c\ln(\Lambda/M)}}\ .
\end{equation}
Taking for example $\Lambda/M\simeq 10$ we may obtain
the upper bound of the
Higgs particle's mass and taking $\Lambda\sim 10^{18}$GeV and
$M\sim 10^3$GeV
the lower bound may be estimated, we have,
\begin{equation}
185/\sqrt{N_c} GeV\leq m_H \leq
720 /\sqrt{N_c} GeV\ ,
\end{equation}
This  result  is
compatible with  the present experimental lower bound
and also lies within the range detectable by future hadron colliders.
Since the Higgs particle's mass is lighter than
1TeV, i.e., the scale signaling the strong
 interaction in the electroweak
symmetry breaking sector, SBVS induced
electroweak symmetry
breaking is "weak", and the
symmetry can be realized linearly in the
Higgs sector. Moreover the low energy
effective theory is $renormalizable$,
all the nonrenormalizable terms are screened by the heavy fermion
mass~\cite{1} ($m^2/M^2$ suppressed).
This is different from the technicolor
interaction (in which the spontaneously broken symmetry
is the chiral symmetry) induced electroweak symmetry breaking.

The correct low energy theory, after integrating out the heavy fermion
fields, should therefore be the effective $O(p^4)$ Lagrangian for Goldstone
fields obtained
in ref.~\cite{1} plus the standard electroweak interaction Lagrangian
of the Higgs field\footnote{
The light Higgs particle brings new
non-decoupling effects. For example
if we further integrate it out
(if it is allowed, i.e., not to be as light as $\sim 2M_W$), at tree level,
there is an additional contribution to $L_1$,
$\delta L_1=v^2/8m_H^2$,
which is the same as the standard model one.
The claim
made in \cite{1} on the difference between the two
$L_1$ terms obtained in the present model and in SM
is therefore incorrect.}. It is also helpful, not to
integrate out fermion
fields completely but firstly down to an arbitrary scale $\mu$ to
study the heavy fermion contributions to the running coupling
constants of the composite Higgs field. We have,
\begin{equation}
\lambda_0(\mu)={N_c\over 8\pi^2}\ln( {\Lambda^2+M^2\over\mu^2+M^2}) \ ,
\end{equation}
\begin{equation}
Z_H(\mu)={N_c\over 4\pi^2}\ln({\Lambda^2 +M^2\over \mu^2+M^2})\ ,
\end{equation}
\begin{equation}
m_H^2(\mu)={N_c\over 2\pi^2}\{ {\pi^2\over G} \Lambda^2-\Lambda^2
+\mu^2 + 3M^2\ln({\Lambda^2+M^2\over\mu^2+M^2})\}\ .
\end{equation}
We see from above expressions that only the high frequency
modes ($\mu >M$) contribute to the wave function
renormalization constant ($Z_H$)
and the bare coupling constant of $\phi^4$ self
interactions ($\lambda_0$).
The low frequency modes only
contribute to  the fine tuning of the Higgs mass.

Once introducing the matter field (quarks and leptons) couplings
 in the same way
as in the SM
we can set up the complete
equivalence between the SM and our model of SBVS, equation (\ref{4f})
\footnote{There is
no difficulty to reproduce the CKM matrix, and even multi Higgs models
in an extended version of four fermion interactions.},
in the $m/M<<1$ limit,
even at the energy scale $E$ much larger
than the electroweak scale as long as $E<<M$,
within the constraints on the Higgs particle's mass.
It is interesting to note that our
model share many low energy properties
of the topcolor model\cite{top},
although we have a very different physical motivation
 from the very beginning.
Our result implies that the Higgs particle's mass
is naturally of the order of the electroweak scale which, if confirmed by
 future experiments,
may therefore
not necessarily be considered  as a support to the topcolor model.

Before conclusion, we would like to stress that it is also appealing
to study the property of heavy vector fermions
in the phenomenology aspects.
As has been pointed out in ref.~\cite{fuji}, the inclusion
of the heavy chiral fermions may violate the  stability of the SM
vacuum. According to  the analysis on the 1--loop effective potential,
the Higgs
mass is therefore forced to become heavy by the appearance
of heavy chiral fermions.
If this is regarded as unnatural in the sense of
perturbative vacuum stability, heavy vector fermions may be
the only reasonable candidates in searching for new
matter constituents of nature.

To conclude, we start from a nonperturbative four fermion interaction
with spontaneously broken
symmetry between heavy fermions in vector
representations of the  symmetry group and derive an asymptotically
renormalizable low energy effective theory,
with a light scalar particle.
This remarkable property of the  decoupling--nondecoupling phase
transition phenomena of SBVS, we believe, is model independent.
To what extent our model will be of realistic importance
when applying to, for example, electroweak physics, may depend on
whether or how can it be read off from a more fundamental
theory since there are restrictions on SBVS\cite{vafa},
if one respects to the gauge interaction as the first principle.
Finally, It is worth emphasizing that there may exist
 the possibility
that  it is vague to say
the Higgs particle is "composite" or "elementary",
 since the two cases
may practically be indistinguishable, as shown
by the above example.

It is my pleasure to thank F. Jegerlehner, M. Locher,
R. Rosenfelder and especially
H. Schlereth for valuable discussions.

\end{document}